\documentclass[aps,prl,reprint,nofootinbib,amsmath,twocolumn,preprintnumbers,superscriptaddress]{revtex4-2}

\usepackage{multirow}
\usepackage[final]{changes}
\usepackage[normalem]{ulem}
\usepackage[hidelinks]{hyperref}
\usepackage[utf8]{inputenc}
\usepackage{graphicx,amsmath,amsfonts,amssymb,slashed}
\usepackage{numprint}
\usepackage{enumitem}      
\usepackage{graphicx,color,amsmath}
\usepackage{xcolor}
\usepackage{bm}
\usepackage{mathtools}
\hypersetup{
    colorlinks,
    linkcolor={blue!50!black},
    citecolor={blue!50!black},
    urlcolor={blue!50!black}
}

\pdfoutput=1
\allowdisplaybreaks
\newcommand{\noise}{\,$\rm e^-_{rms}/pix$}
\newcommand{\sisero}{\mbox{SiSeRO-CCD}}
\newcommand{\siseroamp}{\mbox{SiSeRO}}
\newcommand{\skipper}{\mbox{Skipper-CCD}}

\definecolor{miguel}{rgb}{0.858, 0.188, 0.478}
\begin{document}

\title{Achieving Single-Electron Sensitivity at Enhanced Speed in Fully-Depleted CCDs with Double-Gate MOSFETs}

\author{Miguel Sofo-Haro}
\email{miguelsofoharo@mi.unc.edu.ar}
\affiliation{\normalsize\it Universidad Nacional de C\'ordoba (CNEA and CONICET), C\'ordoba, Argentina}

\author{Kevan Donlon}
\affiliation{\normalsize\it MIT Lincoln Laboratory, Lexington, Massachusetts, USA}

\author{Juan Estrada}
\affiliation{\normalsize\it Fermi National Accelerator Laboratory, Batavia, Illinois, USA}

\author{Steve Holland}
\affiliation{\normalsize\it Lawrence Berkeley National Laboratory, Berkeley, USA}

\author{Farah Fahim}
\affiliation{\normalsize\it Fermi National Accelerator Laboratory, Batavia, Illinois, USA}

\author{Chris Leitz}
\affiliation{\normalsize\it MIT Lincoln Laboratory, Lexington, Massachusetts, USA}

\preprint{...}

\begin{abstract}
We introduce a new output amplifier for \mbox{fully-depleted} thick \mbox{p-channel} CCDs based on \mbox{double-gate} MOSFETs. The charge amplifier is an \mbox{n-type} \mbox{MOSFET} specifically designed and operated to couple the \mbox{fully-depleted} \mbox{CCD} with high charge-transfer efficiency. The junction coupling between the \mbox{CCD} and \mbox{MOSFET} channels has enabled high sensitivity, demonstrating sub-electron readout noise in one pixel charge measurement. We have also demonstrated the non-destructive readout capability of the device. Achieving single-electron and single-photon per pixel counting in the entire \mbox{CCD} pixel array has been made possible through the averaging of a small number of samples. We have demonstrated \mbox{fully-depleted} \mbox{CCD} readout with better performance than the floating diffusion and floating gate amplifiers available today, in both single and multisampling regimes, boasting at least six times the speed of floating gate amplifiers.

\end{abstract}

\maketitle

Since their invention, \mbox{Charge-Coupled} Devices (CCDs) have been the preferred detectors in ground and space telescopes, as well as in laboratory photon imaging applications \cite{boyle1970charge}. Fully-depleted thick CCDs have been developed at Lawrence Berkeley National Laboratory (LBNL) to achieve high quantum efficiency in the red and \mbox{near-infrared} range  
\cite{holland2002overview,holland2003fully,Lesser_2015}. These CCDs are three-phase, p-channel devices, featuring a triple polysilicon gate structure. In order to fully-deplete the thick substrate at a reasonable voltage, they are fabricated in high resistivity \mbox{n-type} silicon \cite{holland2003fully,holland2009device}. Depending on the application, they have been fabricated with thicknesses exceeding \mbox{200\,$\rm \mu m$}, segmented into pixels of \mbox{$\rm 15\times15\,\mu m^2$} and in different array formats \cite{holland2023fully}. To minimize dark current, these CCDs are operated at cryogenic temperatures, typically \mbox{-140\,\textdegree C}. These p-channel CCDs have shown significantly more radiation tolerance than n-channel CCDs, making them more suitable for space-based applications \cite{dawson2008radiation,bebek2002proton}. In the remaining text, we will refer to these CCDs simply as \mbox{FDCCDs}. 

The development of FDCCDs has been driven by their application in detecting and conducting follow-up spectroscopy of high redshift astronomical objects. Examples of instruments utilizing FDCCDs include the Dark Energy Camera (DECam) \cite{dark2016dark,estrada2010focal}, the Baryon Oscillation Spectroscopic Survey (BOSS) \cite{Smee_2013}, and the more recent Dark Energy Spectroscopic Instrument (DESI) \cite{flaugher2014dark}. These FDCCDs, characterized by their low readout noise and considerable active silicon mass, have also found application as competitive direct detectors in dark matter searches \cite{BARRETO2012264,PhysRevLett.123.181802,PhysRevD.94.082006} and the coherent elastic neutrino-nucleus scattering \cite{PhysRevD.91.072001,CONNIE_2019}. For dark matter searches, where active silicon mass needs to be as large as possible, \mbox{FDCCD} are fabricated in \mbox{725\,$\rm \mu m$} thick, \mbox{20\,$\rm k\Omega.cm$} resistivity substrate, and can be fully depleted with \mbox{70\,V} \cite{Cervantes-Vergara_2023}. Today, \mbox{LBNL} and \mbox{MIT} Lincoln Labs (MIT-LL) are leaders in making \mbox{FDCCDs} \cite{howell2006handbook}. Also, relatively thick fully-depleted \mbox{n-channel} CCDs, \mbox{$\sim$100\,$\rm \mu m$}, have been developed by commercial vendors for the Large Synoptic Survey Telescope (LSST) \cite{o2008characterization,gilmore2008lsst}.

Historically, the output amplifier of choice for \mbox{FDCCDs} has been a floating diffusion amplifier (FDA) \cite{holland2003fully}. In the case of the \skipper, the FDA was replaced by a floating gate amplifier (FGA) to enable the non-destructive readout (NDR) of the charge packet, or the so called \mbox{``skipper"} amplifier \cite{fernandez2012sub,holland2023fully}. After averaging the pixel samples, it is possible to achieve a sub-electron readout noise of 0.068\noise, reaching the absolute theoretical limit of silicon of \mbox{1.1\,eV} in energy threshold, allowing single-electron and single-photon sensitivity per pixel \cite{tiffenberg2017single}. The Sub-Electron Noise \skipper\ Experimental Instrument (SENSEI), using a single \skipper\ detector, 675\,$\rm \mu$m thick, has achieved world-leading sensitivity for a large range of \mbox{sub-GeV} dark matter masses \cite{PhysRevLett.121.061803,PhysRevLett.122.161801,PhysRevLett.125.171802}. The \mbox{OSCURA} experiment will lead the search for low-mass dark matter particles using thousands of \skipper\ \cite{aguilar2022oscura,botti2021sub}. \skipper\ are also being explored for terrestrial astronomy \cite{drlica2020characterization}. \mbox{CCDs} are undeniably essential in advancing scientific exploration.

The FGA operates by dumping the pixel charge onto the bulk side of a MOS capacitor in the CCD buried channel to modulate the gate voltage of an output transistor. In addition to the MOS capacitance, several stray capacitances are added to the floating gate (FG) limiting the stage sensitivity. The overall FG capacitance, combined with the resistance of the relatively large poly-silicon connection between the MOS capacitor and the output transistor gate, constrains the time response of the FGA and consequently the maximum pixel readout rate \cite{2023_sisero}. The limited sensitivity of the FG, coupled with the multiple sampling operations, significantly increases the CCD readout time making it a limiting factor in various applications \cite{smart_skipper}.

Another \mbox{FDCCD} with photon counting output and higher pixel readout rate than \skipper is the Electron Multipling CCD (EMCCD). In an extended serial register electrons are accelerated by an electric field and through impact ionization the signal is amplified. They have a limited dynamic range, and have shown degradation due to the gain process, limiting the device lifetime \cite{rauscher2022radiation}. Due to the high radiation tolerance, both \skipper\ and \mbox{EMCCD} are being evaluated as detectors for future space missions in exoplanet search, where single-photon sensitivity is required \cite{rauscher2019radiation}.

In order to overcome the readout noise and speed limitation of FDA and FGA, in this Letter we introduce the first output amplifier for FDCCDs based on a double-gate \mbox{MOSFET}. We will refer to our amplifier as \siseroamp, an abbreviation for Single-Electron Sensitivity Readout, and \sisero\ for the CCD itself. We specifically designed the \siseroamp\, considering its integration and high-voltage compatibility with LBNL fully-depleted \mbox{p-channel} CCDs. For a more comprehensive description of the device design, please refer to \cite{2023_sisero}.

Double-gate MOSFETs were originally proposed by Brewer for the readout of n-channel thin CCDs \cite{brewer1978low}. The amplifier consists of a \mbox{MOSFET} integrated into the CCD buried channel. In this configuration, the CCD channel is junction-coupled to the \mbox{MOSFET} channel, resulting in high-sensitivity modulation of the \mbox{MOSFET} current by the signal charge in the CCD channel \cite{matsunaga1987high,matsunaga1991highly,matsunaga2011low}. Since the signal charge is directly coupled to the \mbox{MOSFET} channel, as reported in \cite{hynecek1999bcd}, this design is the most optimal for any charge detector. Currently, they are being evaluated for the readout of modern \mbox{n-channel} thin CCDs \cite{bautz2019toward,chattopadhyay2021first,chattopadhyay2022first}. In addition to its high sensitivity, another advantage of this amplifier is its NDR capability. Using a similar sense node architecture Depleted Field-Effect Transistors (\mbox{DEPFET}) detectors, which are single \mbox{p-MOSFET} transistors in an \mbox{n-type} depleted thick substrate, have demonstrated a readout noise of \mbox{0.18\noise} after averaging 300 samples in a miniarray of \mbox{4$\times$4} pixels \cite{wolfel2007novel,lutz2016depfet}.

\begin{figure}[ht]
    \centering
    \includegraphics[width=0.35\textwidth]{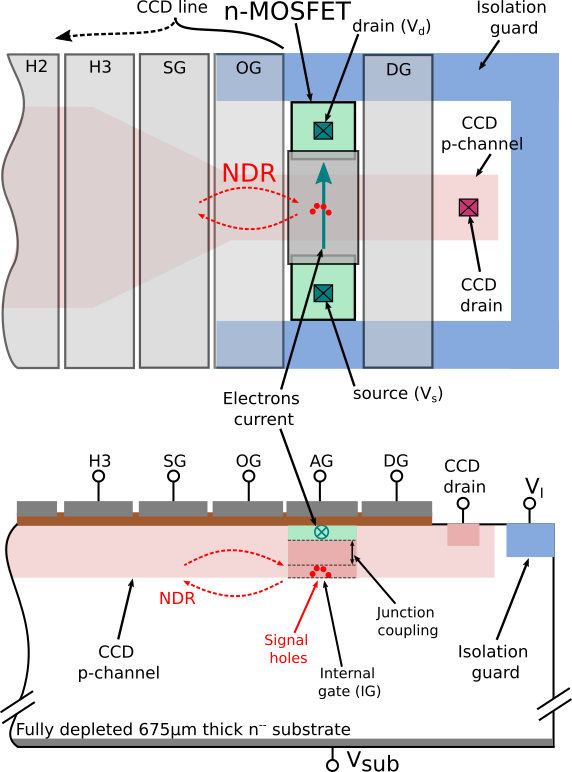}
    \caption{Schematic top view and cross-section diagram of the \sisero\ output amplifier. The \siseroamp\ amplifier consist of a \mbox{n-type} MOSFET transistor integrated into the CCD-line. On the transistor's bulk side, a junction coupling is formed between the transistor \mbox{n-channel} and the CCD \mbox{p-channel}. Due to the direct coupling, high sensitivity and low noise are achieved. Moreover, the charge is sensed non-destructively enabling respective multiple sampling for readout noise reduction.}
       \label{fig:sisero}
\end{figure}

Figure \ref{fig:sisero} illustrates both a top-view and cross-sectional schematic of the \sisero\ output stage. Essentially, it is an \mbox{n-type} \mbox{MOSFET} integrated in the CCD-line, that has the CCD \mbox{p-channel} under the transistor \mbox{n-channel}. This p-n combination creates a junction coupling between both channels. In contrast to the FGA, the charge is directly deposited onto the junction on the bulk side of the transistor, where it directly modulates the junction potential. This change in junction potential modulates the transistor's channel conductance, and depending on the transistor support circuitry, results in either a measurable change in current or voltage. Therefore, the holes deposited into the junction act as an internal type gate (IG) to the common channel of the \mbox{nMOSFET}. By manipulating the gates voltages the charge can be transferred back to the SG to repeat the process and start the acquisition of another sample.   

In order to enable the operation of the \mbox{n-type} \mbox{MOSFET} in the fully-depleted \mbox{n-type} \mbox{CCD} substrate, an isolation guard was designed \cite{2023_sisero}. As is shown in Fig. \ref{fig:sisero} it consist of a \mbox{p-type} implant partially surrounding the output stage. Without the isolation guard, an electron current appears between the \mbox{CCD} backside substrate contact and the \mbox{MOSFET} source/drain terminals, disabling the \mbox{full-depletion} of the \mbox{CCD}. In the \mbox{nMOSFET} bulk-side, a potential well \mbox{(PW)} is formed in the junction between the CCD \mbox{p-channel} and the \mbox{n-type} substrate. The well voltage depends on the transistor biasing point and the acceptor concentration in the junction \cite{2023_sisero}. Another important design aspect was a local p-type implant in the transistor bulk side, depicted as a dark pink region in the bottom diagram of \mbox{Fig.~\ref{fig:sisero}}. This implant creates the PW within the CCD-line well's range after transistor biasing, allowing for the transfer of charge into and out of the IG from the CCD line. As observed in Fig.\ref{fig:sisero}, the CCD-line is tapered in order to reduce the nMOSFET area to achieve a small junction capacitance and increase the sensitivity. It is important to mention that the integration of the \siseroamp\ amplifier does not require modifications of the CCD pixel array; therefore optical characteristics and performance metrics within the array are not affected, such as dark current, quantum efficiency, and radiation tolerance. Moreover, it does not introduce any limitation on the pixel array size. The \sisero\ was fabricated at \mbox{MIT-LL}, using the fabrication process of triple-poly CCDs, in arrays of \mbox{1278$\times$330} pixels in a 725\,$\rm \mu m$ thick substrate, with an output amplifier on each corner for split readout and with a W/L nMOSFET of \mbox{2.5/5\,$\rm \mu m$}. At the short edges of the array, a serial register is placed to shift the charge using three phases (H1, H2, and H3) towards the edges where the \siseroamp\ amplifiers are located.

The \sisero\ was packaged and connected via wire bonding to a \mbox{20\,inches} long flex cable, which serves to transmit signals into and out of the vacuum chamber. Inside the vacuum chamber, the device was cooled down to -140C using a cryocooler. The Low Threshold Acquisition Controller (LTA) was used for the evaluation of the device. Originally developed for \skipper s, the LTA low-noise biasing voltage supplies were adjusted with external hardware to properly bias the \siseroamp. The LTA can be programmed to generate a specific sequence of signals for manipulating the CCD gates. Additionally, it digitizes the video signal and performs in the FPGA the correlated double sampling (CDS) using a double slope integration (DSI) \cite{2021_lta}. 

As previously mentioned, the signal holes in the IG modulate the nMOSFET drain-source current. To convert the current variations into voltage signals, a transimpedance amplifier circuit was implemented using the \mbox{ADA4817} operational amplifier (OA), chosen for its very low input current noise. The OA maintains the \mbox{nMOSFET} drain at a constant potential, thereby preventing any impact of the flex cable's stray capacitance on the readout speed. This biasing and readout circuit ensures the highest speed and performance readout \cite{fischer2003readout}.

\begin{figure}[ht]
    \centering
    \includegraphics[width=0.35\textwidth]{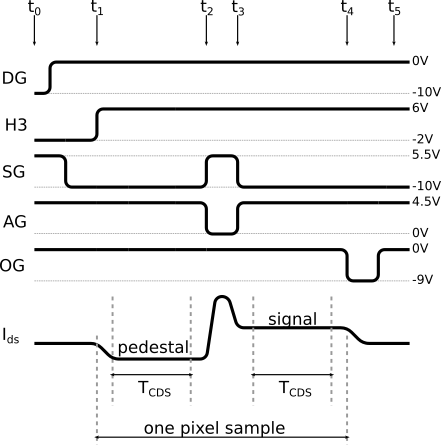}
    \caption{Operating sequence developed for the \sisero\ output stage. At the bottom, there is a representation of the \mbox{nMOSFET} drain-source current ($I_{ds}$). Between $t_1$ and $t_2$, the IG is empty of signal holes and the pedestal level integration takes place for $T_{CDS}$. Between $t_2$ and $t_3$, the SG is pulled up and the \mbox{nMOSFET} is turned off to ensure the transfer of the signal holes into the IG. The signal level integration takes place between $t_3$ and $t_4$. At $t_4$, OG is pulled down to transfer the charge back into the SG to repeat the process to take another pixel sample.} 
 
    \label{fig:sequencer}
     
\end{figure}

To enable multi sampling operation and achieve the single-electron sensitivity regime, a specific operating sequence had to be developed for the \sisero. Figure~\ref{fig:sequencer} depicts the developed operating sequence. It has been developed as follows: At $t_0$, the holes are drained from the IG to the CCD drain by lowering the dump gate (DG). At $t_1$, the horizontal phase H3 is raised to transfer the pixel's holes into the summing gate (SG) well. During this time, the pedestal integration takes place for a duration of $T_{CDS}$. At $t_2$, the SG is raised to transfer the holes over the output gate (OG) into the IG. The well voltage of the IG depends on the \mbox{nMOSFET} biasing voltages. Higher drain and gate voltages would increase the PW voltage \cite{2023_sisero}. Consequently, the charge transfer from the SG into the IG is largely influenced by the biasing point. To mitigate this dependence, we strategically turn off the \mbox{nMOSFET} (by setting AG to 0V) during this operational phase. Together with the additional local \mbox{p-type} implant, this ensures that the IG well voltage is lower than the OG well voltage, creating a sufficient voltage gradient to fully transfer the signal holes from the SG into the IG. At $t_3$, with the holes in the IG, both SG and AG are restored to initiate the integration of the signal level. We observed that the SG and OG are strongly capacitively coupled to AG. As a result, the voltage levels on both gates can modify drain-source current ($I_{ds}$) and consequently the gain. Therefore, they are maintained in the same state during both the pedestal and signal integration periods to ensure equal gain. The pixel charge value is determined by the difference between the signal and pedestal levels. At $t_4$, the OG is set low to transfer the holes back into the SG, repeating the process for taking another sample of the pixel charge. Figure~\ref{fig:sequencer} provides the high and low voltage levels of each signal as a function of time. 

The isolation guard was biased with \mbox{8\,V}, which was predetermined through computer simulations \cite{2023_sisero}. The \mbox{nMOSFET} was operated at a constant biasing point. In order to minimize impact ionization noise, the drain-source voltage ($V_{ds}$) was biased to \mbox{1\,V} with the source at \mbox{0V} \cite{2023_sisero}. The \mbox{nMOSFET} is operating in enhancement mode, with a threshold voltage of \mbox{2.4\,V}. A maximum sensitivity of \mbox{1.54\,$nA/e^-$} was achieved with $V_{gs}=4.5V$. At this operating point, the \mbox{source-drain} current was \mbox{70\,$\rm \mu A$}. This is the highest sensitivity ever achieved with a \mbox{double-gate} \mbox{MOSFET}. Typically DEPFETs has a sensitivity of \mbox{$\sim$0.3\,$nA/e^-$} \cite{wolfel2007novel}. In the case of thin n-type CCD readout, 0.7\,$nA/e^-$ have been reported in \cite{chattopadhyay2021first}.
  
\begin{figure}[t]
    \centering
    \includegraphics[width=0.45\textwidth]{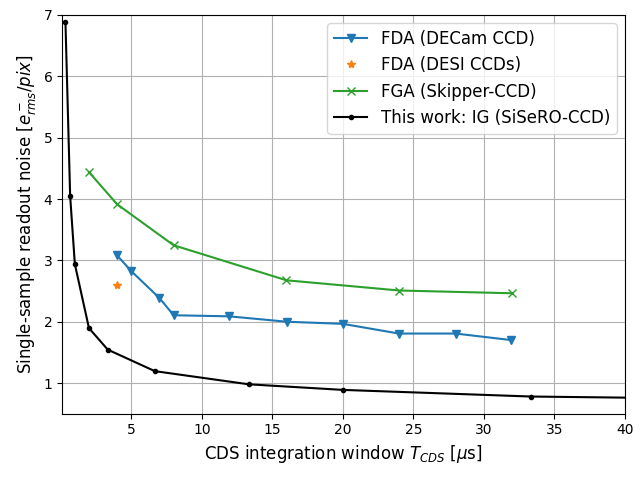}
    \caption{Single-sample readout noise versus the correlated-double sampling (CDS) integration time $T_{CDS}$. As observed, the \sisero\ consistently achieves lower readout noise throughout the entire integration time range, highlighting the advantage of its internal gate (IG) sensing structure over fully-depleted thick CCDs with FDA \cite{holland2003fully,holland2023fully} or FGA \cite{2021_lta} output structures.}
    \label{fig:noise_vs_T}
\end{figure}

Images of cosmic-ray muons and x-rays were acquired with the \sisero. The achieved single-sample pixel readout noise at different CDS integration times ($T_{CDS}$) is shown in Fig.~\ref{fig:noise_vs_T}. A readout noise of 2.72\noise\ is achieved for a $T_{CDS}$ of \mbox{1\,$\rm \mu s$}. Over \mbox{20\,$\rm \mu s$}, the noise decreases down to 1\noise\ and can reach a floor of 0.74\noise\ at \mbox{100\,$\rm \mu s$}. For comparison, results achieved in FDCCDs with FDA and FGA output stages are also presented. The FDA results are obtained from the Dark Energy Camera (\mbox{DECam}) CCDs \cite{estrada2010focal}, as extracted from \cite{holland2003fully}. Additionally, the result from the Dark Energy Spectroscopic Instrument (DESI) CCDs is included \cite{holland2023fully}, that have an optimized FDA output structure \cite{bebek2017status,holland2014technology}. The FGA case corresponds to the \skipper\, as extracted from \cite{2021_lta}. As observed, the \siseroamp\ achieves a lower readout noise over the entire integration integration time range compared to the other output structures.

\begin{figure}[ht]
\centering
\includegraphics[width=0.45\textwidth]{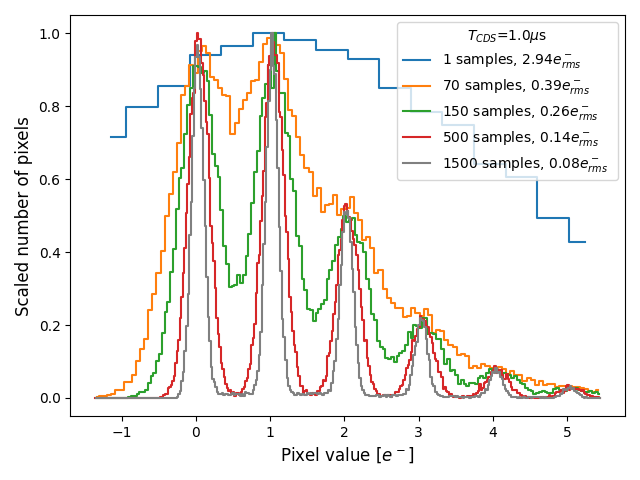}
\includegraphics[width=0.45\textwidth]{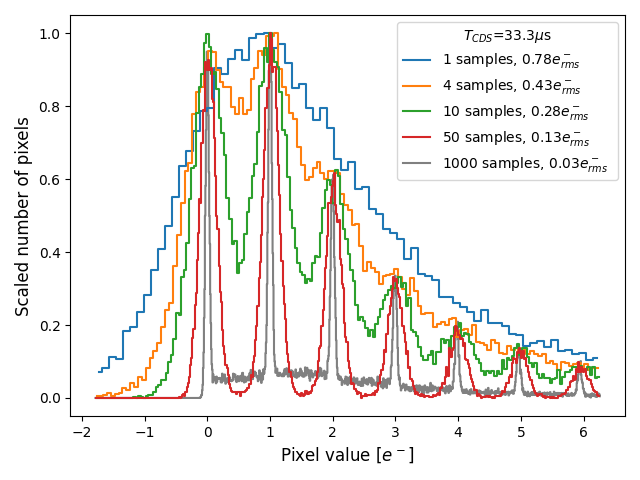}
\caption{Single pixel distribution of an image acquired with $T_{CDS}$ of  \mbox{1.0\,$\rm \mu s$} (top) and \mbox{33.3\,$\rm \mu s$} (bottom). The charge quantization starts to be revealed as the number of averaged samples increases, and \mbox{single-electron} charge resolution is achieved in both cases with the \sisero.}
\label{fig:peaks}
\end{figure}

Figure~\ref{fig:peaks} shows the pixel charge distribution of an image acquired with $T_{CDS}$ of \mbox{1\,$\rm \mu s$} and \mbox{33.3\,$\rm \mu s$}, after averaging different number of pixel samples ($N_{spl}$). As observed, \mbox{single-electron} resolution is achieved in both cases, and the unprecedented readout noise of 0.03\noise\ is achieved for \mbox{33.3\,$\rm \mu s$} and 1000 samples. Charge quantization has been also observed up to 600\,$e^{-}$, showing \mbox{single-electron} sensitivity in a wide dynamic range. As a drawback, for long pixel readout times, that is proportional to $N_{spl}$ and $T_{CDS}$, impact ionization holes produced by the nMOSFET electron current, can fall into the internal gate during the sampling operation \cite{2023_sisero}. This affect the measurement, as can be observed in the case of 33\,$\rm \mu s$ and 1000 samples of Fig.~\ref{fig:peaks}, resulting in non-zero pixel count between the discrete electron levels. This is due to a single carrier from an impact ionization event collecting in the sense node at some point during the readout.

\begin{figure}
    \centering
    \includegraphics[width=0.45\textwidth]{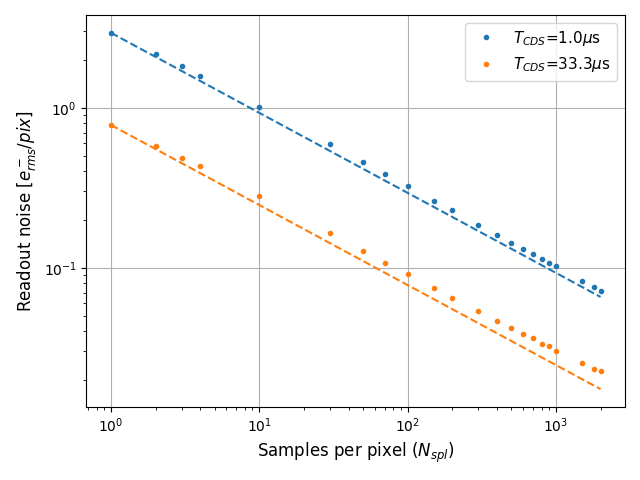}
    \caption{\sisero\ readout noise versus numbers of averaged samples per pixel ($N_{spl}$). The dashed lines represent the $1/\sqrt{N_{spl}}$ projection based on single sample readout noise, as expected from uncorrelated samples.}
    \label{fig:noise_vs_nsamp}
\end{figure}

Fig.~\ref{fig:noise_vs_T} shows the achieved readout noise after averaging different number of samples for different $T_{CDS}$. If the pixel samples are affected by uncorrelated noise, the readout noise decreases in proportion to the square root of the number of averaged samples \cite{tiffenberg2017single}. As observed in \mbox{Fig.~\ref{fig:noise_vs_nsamp}}, this is the case for $T_{CDS}$ of \mbox{1\,$\rm \mu s$}. As discussed in \cite{fernandez2012sub}, with an increase in pixel readout time, the transistor flicker noise introduces correlation among the samples. This correlation effect is observable in \mbox{Fig.~\ref{fig:noise_vs_nsamp}} for longer $T_{CDS}$, where the achieved readout noise deviates from the behavior exhibited by uncorrelated samples. The minimum achieved readout noise has been 0.021\noise\, for \mbox{86.7\,$\rm \mu s$} of $T_{CDS}$ and 2000 samples.

\begin{figure}
    \centering
    \includegraphics[width=0.45\textwidth]{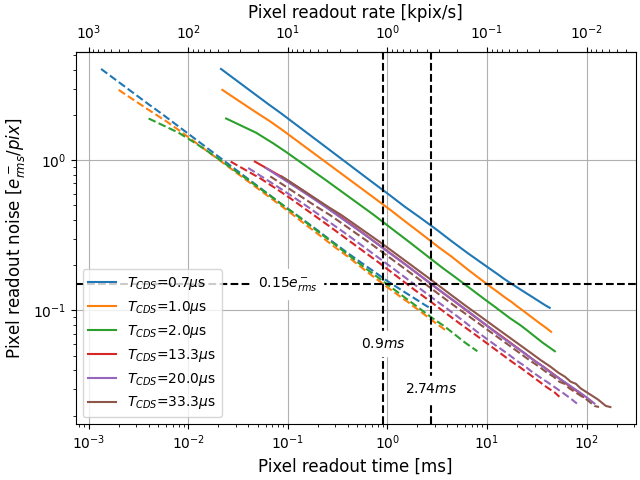}
    \caption{\sisero\ pixel readout time. A readout noise of \mbox{0.15\noise} is achieved in a pixel readout time of \mbox{2.74\,ms} \mbox{(365 pixels/sec)} using a $T_{CDS}$ of \mbox{13.3\,$\rm \mu s$}. The minimum possible readout time, considering two times $T_{CDS}$ as the pixel sample readout time (dashed lines), is \mbox{0.9\,ms} \mbox{(1.1\,kpixels/sec)} for a $T_{CDS}$ of \mbox{1\,$\rm \mu s$}.}
    \label{fig:sisero_speed}
\end{figure}

With the LTA readout system, \mbox{20\,$\rm \mu$s} for signal manipulation was required per sample during the multi-sampling operation. Therefore the pixel readout time is given by \mbox{$(2 \times T_{CDS}+20\mu s)\times N_{spl}$}. \mbox{Fig.~\ref{fig:sisero_speed}} shows the achieved pixel readout noise versus the pixel readout time for different values integration time. As observed, the minimum readout time is achieved with a $T_{CDS}$ of \mbox{13.3\,$\rm \mu s$}, reaching \mbox{0.15\noise} in \mbox{2.4\,ms} \mbox{(365\,pixels/sec)}, which is six times faster than the Skipper-CCD \cite{tiffenberg2017single}. In \mbox{Fig.~\ref{fig:sisero_speed}}, the dashed lines represent the readout speed limit that could be reached by reducing the readout controller signal manipulation time down to the ideal \mbox{0\,$\rm \mu s$}. As observed, the minimum readout time is achieved with a $T_{CDS}$ of \mbox{1\,$\rm \mu s$}, resulting in a readout noise of \mbox{3\noise} at \mbox{500\,kpixels/sec} and \mbox{0.15\noise} at \mbox{1.1\,kpixels/sec}.

With further optimization the device performance can be improved. One possible design optimization is to develop a buried-channel \mbox{nMOSFET}, implanting the transistor n-channel as described in \cite{2023_sisero}, in order to further reduce the noise floor of 0.74\noise\ for long integration times where flicker noise is dominant. Another design optimization can be to reduce the area of the p-type local implant in order to enhance the coupling between the channels, and to reduce the coupling of the IG to the \mbox{nMOSFET} source and drain diffusion where a loss of sensitivity occurs \cite{2023_sisero}. It is worth note that already developed techniques for the \skipper\ are also applicable for the \sisero. For example, smart readout of the array, where only a subset of the array is readout taking several samples for noise reduction, can be combined with the \sisero\ to boost the array readout speed \cite{smart_skipper}. Multiplexed readout architectures with analog averaging developed for the readout of large arrays are also suitable for this detector \cite{skipper_mux}. 

In summary, we have demonstrated that double-gate MOSFETs serve as highly sensitive charge amplifiers for fully-depleted CCDs, setting new records in readout noise and speed while achieving single-electron sensitivity. An unprecedented level of sensitivity and low readout noise at shorter pixel readout times, compared to traditional FDA and FGA outputs, is attainable. Furthermore, it demonstrates single-electron sensitivity even after averaging relatively few samples. In future work, we will continue to improve performance by optimizing the device itself, the readout electronics, and characterizing its performance in specific application cases. 

\bibliography{bibdoi}
\end{document}